\documentclass{aa}  

\newcommand{\mmax}{m_{\mathrm{max}}}
\newcommand{\mob}{M_{\mathrm{OB}}}

\newcommand{\rtid}{R_{\mathrm{tid}}}
\newcommand{\rh}{R_{\mathrm{h}}}
\newcommand{\mrh}{M_{\mathrm{rh}}}
\newcommand{\rc}{R_{\mathrm{c}}}
\newcommand{\mtid}{M_{\mathrm{tid}}}
\newcommand{\ftid}{f_{\mathrm{tid}}}
\newcommand{\mbh}{M_{\mathrm{BH}}}
\newcommand{\nrh}{N_{\mathrm{rh}}}
\newcommand{\abbh}{a_{\mathrm{BBH}}}
\newcommand{\mave}{\langle m_{\mathrm{s}} \rangle}
\newcommand{\ns}{N_{\mathrm{s}}}
\newcommand{\ntid}{N_{\mathrm{tid}}}
\newcommand{\mout}{M_{\mathrm{out}}}
\newcommand{\nout}{N_{\mathrm{out}}}
\newcommand{\trh}{T_{\mathrm{rh}}}

\usepackage{graphicx}
\usepackage{txfonts}
\usepackage{color}
\usepackage{listings}
\begin{document}

   \title{The impact of massive stars and black holes on the fate of open star clusters and their tidal streams}

   \titlerunning{The impact of massive stars on open star clusters}

   \author{Long Wang
          \inst{1,2}
          \and
          Tereza Jerabkova\inst{3}
          }

   \institute{Department of Astronomy, School of Science, The University of Tokyo, 7-3-1 Hongo, Bunkyo-ku, Tokyo, 113-0033, Japan \\
              \email{long.wang@astron.s.u-tokyo.ac.jp}
         \and
         RIKEN Center for Computational Science, 7-1-26 Minatojima-minami-machi, Chuo-ku, Kobe, Hyogo 648-0047, Japan
         \and
             European Space Agency (ESA), European Space Research and Technology Centre (ESTEC), Keplerlaan 1, 2201 AZ Noordwijk, The Netherlands,\email{Tereza.Jerabkova@esa.int}
             }

   \date{Received June 15, 2021; accepted --, --}

 
  \abstract
   {To investigate how the content of massive OB stars affects the long-term evolution of young open clusters and their tidal streams, and how such an effect influences the constraint of initial conditions by looking at the present-day observations.}
   {OB stars are typically in binaries, have a strong wind mass loss during the first few Myr, and many become black holes. These affect the dynamical evolution of an open star cluster and impact its dissolution in a given Galactic potential. 
   We investigate the correlation between the mass of OB stars and the observational properties of open clusters. 
   Hyades-like star clusters are well represented in the Solar neighborhood and thus allow comparisons with observational data. }
   {We perform a large  number of star-by-star numerical $N$-body simulations of Hyades-like star clusters by using the high-performance $N$-body code \textsc{petar} combined with \textsc{galpy}.   
   }
   {
   We find that OB stars and black holes have a major effect on star cluster evolution. 
   Star clusters with the same initial conditions, but  a different initial content of OB stars, follow very different evolutionary paths. 
   Thus,  the initial total mass and radius of an observed star cluster cannot be unambiguously determined unless the initial content of OB stars is known.
   We show that the stellar counts in the corresponding tidal tails, that can be identified in the Gaia data,  help to resolve this issues. 
   We thus emphasise the importance of exploring not only star-clusters, but also their corresponding tidal tails. 
   These findings are relevant for studies of the formation of massive stars. 
   }
   {}


   \maketitle
%

\section{Introduction}

Young open star clusters are important stellar systems for understanding star formation and stellar evolution. 
Due to the stellar and dynamical evolution, star clusters are evaporating their stars and are in the process of becoming part of the Galactic field stellar population. 
Thus, for many star clusters a large fraction of their stars are not anymore located in the clusters themselves, but in the large-scaled tidal tails shaped by the Galactic tides and internal cluster processes \citep{Baumgardt2003, HH03,CR06,  Kuepper2008, Kuepper+10, Dinnbier2020a,Dinnbier2020b,Jerabkova2021}.

The Gaia survey, with its currently most up to date third data release EDR3 \citep{GaiaeDR32021}, provides rich astrometric and photometric data allowing to study kinematic properties of stars in the solar neighborhood in unprecedented detail.
A new era comes as more and more new open clusters \citep[e.g.][]{Cantat2018,Cantat2019, Ginard2018, Ginard2020, Sim2019, Liu2019, Ferreira2020, He2021}, large scale co-eval relic filaments \citep{Jerabkova2019, Beccari2020} and extended stellar streams \citep{Kounkel2019}  have been discovered recently using Gaia.
With the Gaia data, for the first time, it has become possible to extract tidal tails of open star clusters, a challenging task due to  the difficulty to distinguish  tail stars from the Galactic field population
\citep[e.g.][]{Roeser2019a,Roeser2019b,Furnkranz2019,Meingast2019,Tang2019,Zhang2020, Bhattacharya2021, Jerabkova2021}. 
The kinematic properties of Galactic young star clusters \citep[e.g.][]{Kuhn2019,Monteiro2019,Monteiro2020,Zhong2020,Angelo2021,Dias2021,Godoy-Rivera2021} and their extended structures \citep[e.g.][]{Pang2020,Pang2021,Meingast2021} have also been significantly improved.

With data on the tidal tails of open clusters becoming available, the initial conditions of these can be better constrained. 
These in turn constrain star formation theories: Do all stars form in initially gravitationally bound clusters? 
What are the radii and masses of these, and is the stellar and binary population in these always the same?
Inferring the initial conditions of star clusters constitutes an important astrophysical problem, because the observed coeval stellar population can then be matched to the physical conditions of its birth.
However, star cluster formation starting from giant molecular clouds involves a multi-scale complexity. 
The formation of a star cluster is not an isolated event, but is embedded in a large-scale distribution of molecular could.
For example, the Orion molecular clouds has a length scale of hundreds of pc where several star-forming regions exist. \citep[e.g.][]{Bally1987,Genzel1989,Carperter2000,Dame2001,Kounkel2018,Jerabkova2019,Beccari2020}.

Meanwhile, in the dense region where star clusters form, the wind, the radiation and the supernovae feedback from massive OB stars significantly affects the star formation.
They can remove the gas from the cluster-forming region in a short timescale ($\approx1\,$Myr) \citep[e.g.][]{KAH, Goodwin2006,Baumgardt2007, DW20, Semadeni20, Fujii2021b}.
If the star formation efficiency is low, the gas expulsion can significantly change the gravitational potential and even cause the immediate destruction of the system.
The feedback can also affect the surrounding large-scale molecular cloud, such as the formation of the Orion–Eridanus Superbubble \citep[e.g.,][]{Blaauw1964,Brown1994,Madsen2006,Odell1967,Odell2011,Ochsendorf2015,Kounkel2020,Grossschedl2021}.

It is challenging to fully understand this complexity in star formation. 
However, during the post-gas-expulsion phase, if a star cluster was still bound, the stellar dynamics would re-virialise the cluster and would result in a roughly spherical stellar system as observed today. 
In this work, we strengthen that the massive OB stars can also significantly influence the long-term dynamical evolution of the post-gas-expulsion star clusters.
The wind mass loss of OB stars after gas expulsion continues to reduce the gravitational potential of the star cluster.
Meanwhile, the death of OB stars can leave compact objects like black holes (BHs).
By being much more massive than stars, they are the most likely candidates to end up in dynamically formed binaries. 
The interaction between these binaries and surrounding stars is a heating mechanism that influences the long-term evolution of the system \citep[e.g.][]{Binney1987,Spitzer1987,Mackey+08, Breen2013}.

The  number of OB stars formed is related to the fundamental nature of the initial mass function (IMF) in a star-forming region \citep{Kroupa+13}.
If star formation were to be stochastic, the number of OB stars in a gas cloud would be described by randomly sampling from the IMF.
Thus, even if a group of low-mass open clusters  would have the same initial properties (i.e. the same masses, radii), they would show noticeable differences after hundreds of Myr of dynamical evolution due to the large variation of the number of OB stars. 
In contrast, if star formation were to be highly self-regulated, a strong correlation of the number of OB stars and the total mass of the cluster-forming gas cloud would result, and the stochastic effect  would be significantly suppressed \citep[e.g.,][]{Weidner2006,Weidner2013,Kroupa+13, Yan2017}.

By numerically modelling the evolution of Hyades-like open clusters, we can quantify the cluster bulk properties (mass, radius) as a function of time. 
For low-mass open star clusters which only contain thousands of stars or fewer, the stochastic effect from the randomization of initial masses, positions and velocities of stars cannot be ignored. 
The evolution of cluster, such as the density, morphology and the distribution of escaping stars, diverge increasingly with time due to the inherent chaotic nature of the system \citep{Heggie88, Goodman+93, Heggie96}. 
By allowing, for a model of fixed initial mass, the IMF, positions and velocities of stars to be randomly sampled, we study the degeneracy of initial configurations of post-gas-expulsion, expanded re-virialised Hyades-like open clusters (the most  comprehensively observed on) in order to constrain the maximum possible variation of such initial conditions that yield a comparable present-day configuration.
The maximum variation is obtained by treating the IMF as a probabilistic distribution function rather than an optimally sampled distribution function \citep{Kroupa+13}. 
We point out for the first time that by combining the present-day properties of the open cluster (its mass, radius, stellar population) with the population of stars in its tidal tails and their extend, the degeneracy can be broken: the post-gas-expulsion initial conditions for an observed open cluster can be inferred uniquely if its astrophysical age, its bulk properties and the stellar population in its tidal tails are known. 
In particular, the initial content of OB stars can be determined. 
Such constraints can then, in the future, be linked to the pre-gas-expulsion birth conditions by applying hydrodynamical modelling of the forming embedded cluster \citep[e.g.][]{Hirai2021,Fujii2021a,Fujii2021b}.


First, in Section~\ref{sec:method}, we describe the numerical $N$-body code, \textsc{petar} equipped with \textsc{galpy}, and the initial condition for modelling the formation and evolution of a Hyades-like star cluster and its tidal stream.
Then, in Section~\ref{sec:result}, we analyze the observational features and dynamical evolution of our models in detail. Especially, we focus on how the initial mass of OB stars affects the present-day properties of the cluster and its tidal stream.
Finally, we summarize and discuss our findings in Section~\ref{sec:conclusion}.

\section{Method}
\label{sec:method}

\subsection{The $N$-body code \textsc{petar}}
In this work, we use the $N$-body code \textsc{petar} \citep{Wang2020c} to perform the numerical simulations of the star clusters.
\textsc{petar} is a high-performance $N$-body code that combines three algorithms: 
\begin{itemize}
\item the Barnes-Hut tree \citep{Barnes1986} for the long-range interactions; 
\item the fourth-order Hermite integrator with block time steps for middle-range interactions \citep[e.g.,][]{Aarseth2003}; 
\item and the slow-down algorithmic regularization method \citep[SDAR;][]{Wang2020b} for short-range interactions inside multiple systems, such as hyperbolic encounters, binaries and hierarchical few-body systems.
\end{itemize}
The three algorithms are combined via the particle-tree particle-particle method \citep{Oshino2011}.
The framework for developing parallel particle simulation codes (\textsc{fdps}) is used to achieve a high performance with the multi-process parallel computing \citep{Iwasawa2016,Iwasawa2020,Namekata2018}.

Although the Barnes-Hut tree introduces an approximation of the force calculation to reduce the computational cost for the massive stellar systems with $>10^5$ stars, \textsc{petar} can also well simulate low-mass systems when the accuracy parameters are set properly.
The high performance and the low computational cost allow us to carry out a large ensemble of $N$-body simulations of low-mass clusters. 

\subsection{The stellar evolution package \textsc{sse/bse}}

The recently updated version of single and binary stellar evolution packages, \textsc{sse} and \textsc{bse}, are used to simulate the wind mass loss, the type changes of stars, the mass transfer and the mergers of binaries \citep{Hurley2000,Hurley2002,Banerjee2020}.
These population synthesis codes use the semi-empirical stellar wind prescriptions from \cite{Belczynski2010}.
For the formation of compact objects, we adopt the ``rapid'' supernova model for the remnant formation and material fallback from \cite{Fryer2012}, along with the pulsation pair-instability supernova \citep[PPSN;][]{Belczynski2016}.
We apply the solar metallicity ($Z=0.02$) in this study.

\subsection{The galactic potential package \textsc{galpy}}

The \textsc{galpy} code \citep{Bovy2015} has been implemented as a plugin to \textsc{petar} via the c programme language interface. 
The \textsc{petar-galpy} combination can simulate the formation and evolution of massive star clusters with their tidal streams in a variety types of galactic potentials (see Appendix~\ref{sec:galpy}).
In addition, we also implement the \textsc{petar}-Gaia Python tool for the data analysis and the comparison with observational data (see Appendix~\ref{sec:obs}).

The Python interface of \textsc{galpy} provides more functions than the c interface.
However, the latter has a much better computing performance.
In this work, using \textsc{petar} we can efficiently generate a large amount of simulations for dynamically evolving open star clusters in the Milky-way Potential.
One simulation listed in the Section~\ref{sec:init} needs a computing wall-clock time of 7~min to 4~hours with one CPU core.
The 4500 models in this study cost about 4 days computing to finish by using a desktop computer with a AMD Ryzen Threadripper 3990x CPU (64 cores). 

In our models, we setup the Galactic potential by using the MWPotential2014 from \textsc{galpy}.
This is a simplified static Galactic potential model including a bulge, a disk and a dark-matter halo.
Since exactly reproducing the observed structure along the tidal stream of Hyades is not the purpose of this work, this approximated potential is sufficient for our study.
The parameters of MWPotential2014 are listed as following:
  \[
      \begin{array}{lp{0.8\linewidth}}
         Bulge  & a power-law density profile with an exponentially cut-off      \\
                & -- power-law exponent: -1.8 \\
                & -- cut-off radius: 1.9 kpc                    \\
                & -- mass: $5\times 10^9 M_\odot$ \\
         Disk &  \cite{Miyamoto1975}  disk \\ 
              & -- scale length: 3.0~kpc\\
              & -- scale height: 280~pc \\
              & -- mass: $6.8 \times 10^{10} M_\odot$ \\
         Halo & NFW profile \citep{NFW1995} \\
              & -- scale radius: 16~kpc \\
              & -- local dark-matter density: 0.008 $M_\odot pc^{-3}$ \\
      \end{array}
   \]
\noindent
The MWPotential2014 assumes a solar distance to the Galactic center to be 8 kpc and the solar velocity to be 220 km s$^{-1}$.

\subsection{Initial conditions}
\label{sec:init}

\begin{figure}[htbp]
    \centering
    \includegraphics[width=\columnwidth]{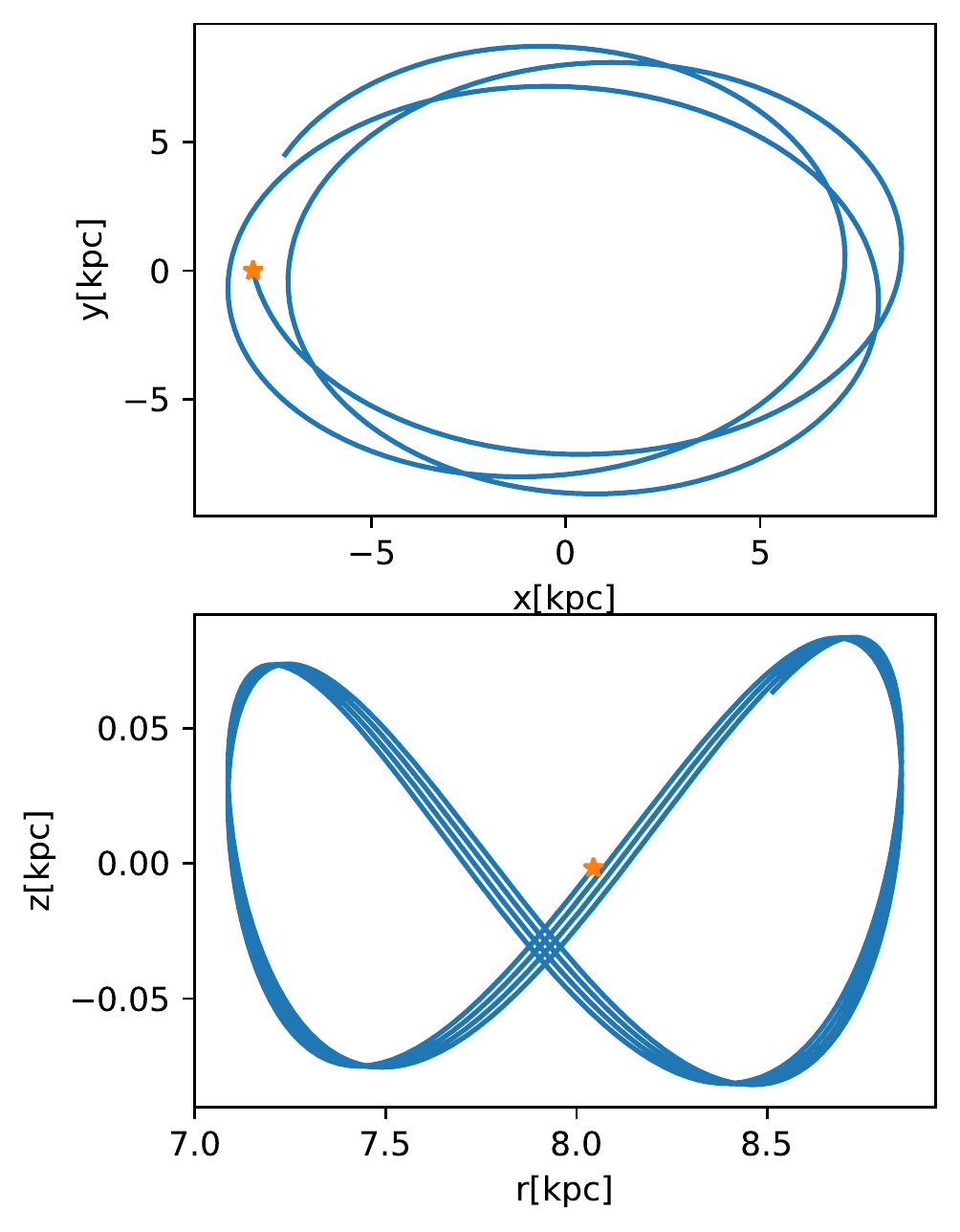}
    \caption{The orbit of Hyades-like cluster in the Galactocentric reference frame calculated by using \textsc{galpy}. The upper panel shows the orbit in the $x$-$y$ plane and the lower panel shows the orbit in the $r$-$z$ plane, where $r$ is the radial distance projected in the x-y plane. The symbol of star represents the present-day position.}
    \label{fig:orbit}%
\end{figure}

\subsubsection{The real Hyades cluster}
\label{sec:orbit}
The Hyades is the closest star cluster to the Sun,  with a distance of 45 pc. 
\cite{Roeser+11, Roeser2019b} constrained the Hyades' half-mass and the tidal radii to be 4.1 pc  
and 9 pc (up to 10.4 pc), respectively, with the total stellar mass of the Hyades being 435 $M_{\odot}$ (up to 509 $M_{\odot}$) and the bound mass being 275 $M_{\odot}$ (up to 322 $M_{\odot}$). The upper estimated are correction based on assumptions about unresolved binaries, for more details see  \cite{Roeser+11}. See also \cite{Reino2018} for constraints on the Hyades cluster membership and velocity dispersion and relevant discussion in \cite{Jerabkova2021}.

To find the initial position and velocity of the Hyades cluster in the Galaxy, 
we reverse the velocity of the center of Hyades at the present day and integrate back 648 Myr by using the time-symmetric integrator in \textsc{galpy} to find the initial position (see Figure~\ref{fig:orbit}).
Then, we reverse the velocity 648 Myr ago to obtain the initial velocity.
Since the MWPotential2014 is static, if we start from this initial coordinate and integrate forwards to the present day, we can recover the correct position and velocity with a small numerical error.
The evaluated initial and the observed present-day coordinates of the Hyades in the International Celestial Reference System (ICRS) frame are listed in Table~\ref{tab:init} \citep[see][for more details]{Jerabkova2021, Gaia_Hyades}.

The stellar wind mass loss and supernovae can break the conservation of the momentum.
Thus, in the $N$-body simulation using \textsc{petar}, the center of the star cluster may not exactly follow the orbit obtained from the result using \textsc{galpy}.
But the offset at the present day is small so that it does not affect our analysis.
We can also correct the offset by rotating the coordinates of the stellar system in the Galactrocentric frame\footnote{based on \textsc{coordinates.basecoordinateframe} from the \textsc{astropy} Python module \citep[see][]{astropy2013,astropy2018}}. 

   \begin{table}
      \caption[]{The initial and present-day coordinates of Hyades}
        \label{tab:init}
         \begin{tabular}{lll}
            \hline
            \noalign{\smallskip}
            Parameter & 648 Myr ago & present-day \\
            \noalign{\smallskip}
            \hline
            \noalign{\smallskip}
            RA[$\deg$] & 308.488  & 67.985 \\
            Dec[$\deg$] & 41.217 &  17.012\\
            Distance[pc] & 4.575 & 47.501\\
            pm(RA,cos(Dec))[mas yr$^{-1}$] & -0.904 & 101.005\\
            pm(Dec) [mas yr$^{-1}$] & -0.740 & -28.490\\
            radial velocity [km s$^{-1}$] & 431.31 & 39.96 \\
            \noalign{\smallskip}
            \hline
         \end{tabular}
         \tablefoot{The evaluated coordinate of the Hyades 648 Myr ago by using \textsc{galpy} and the observed present-day one in the ICRS frame \citep[see][for more details]{Jerabkova2021, Gaia_Hyades}}
   \end{table}

\subsubsection{The models}

\begin{figure*}[htbp]
    \centering
    \includegraphics[width=0.9\textwidth]{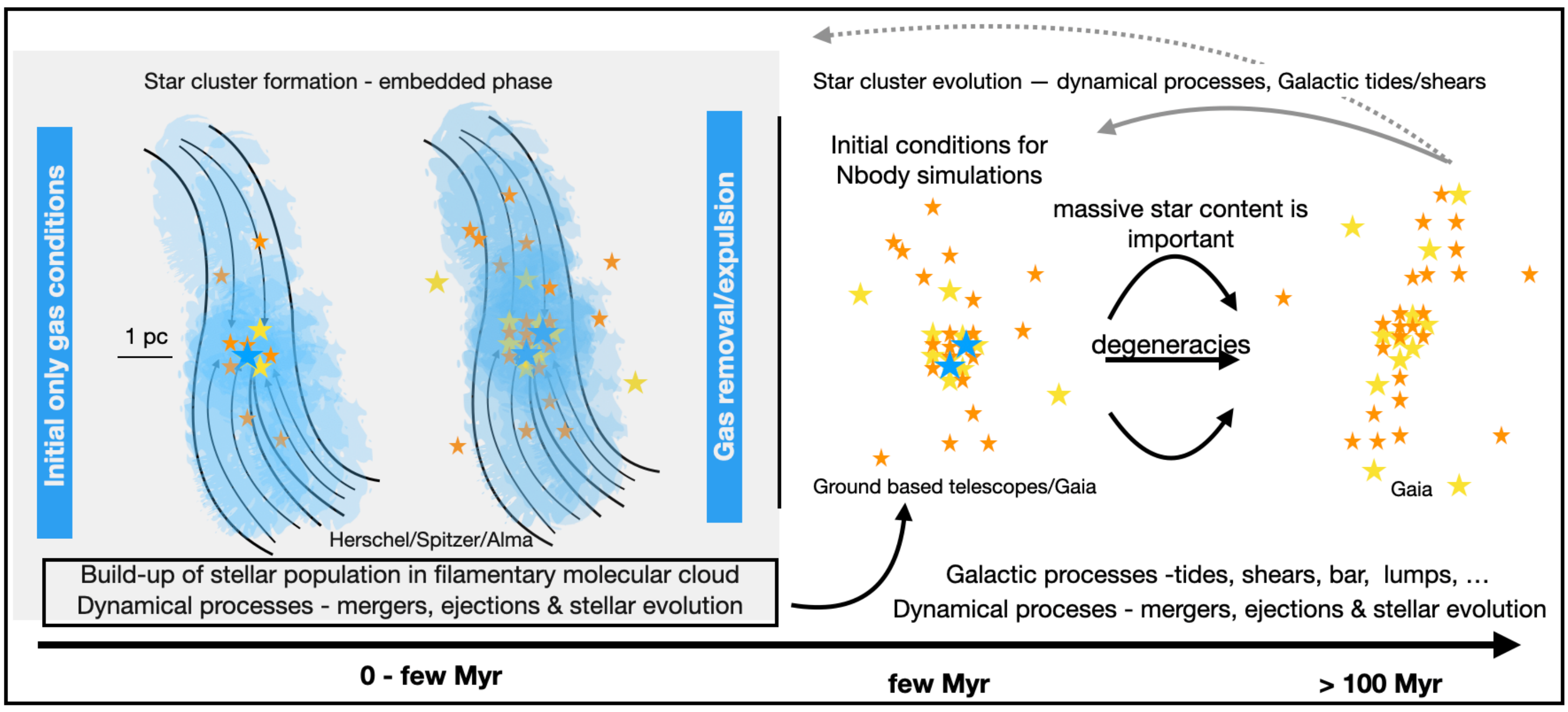}
    \caption{Sketch of the formation and evolution of a star cluster (from left to right). The left: the gas embedded phase where star formation occurred along the filaments in a few Myr, which were influenced by the gas flow in the large scale, the OB feedback and the stellar dynamics. The right: after the gas expulsion, the bound star cluster re-virialised to a roughly spherical system, which is the initial condition in our models. With the three black solid arrows we show that there are a number of evolutionary tracks a star cluster can take. In our case we focus on how the content of massive stars can affect the evolution of the cluster with otherwise identical initial conditions. We show that while this causes degeneracy making it impossible to uniquely infer the initial conditions of an observed cluster, the properties of its tidal tails help to break the degeneracy and thus constrains the initial conditions, as marked by the gray solid arrow. The ultimate question (indicated by the gray dotted arrow) is then what constraints can be placed on the gas-embedded phase of the star cluster.   }
    \label{fig:sketch}%
\end{figure*}

Fig.~\ref{fig:sketch} shows a sketch describing the phase transition for the evolution of a star cluster.
The gray-shaded panel shows the gas-embedded phase involving the following general steps: 
i) initial convergent gas flow towards a density maximum, 
ii) on-set of star-formation along the filaments and predominantly at their intersection comprising the bulk of the later stellar body of the embedded star cluster, 
iii) star cluster formation being affected by dynamical processes (mergers, ejections) and stellar feedback as exemplified by the Orion Nebula Cluster \citep{Kroupa+18}, 
iv) gas expulsion which gradually begins with the on-set of star formation and results in a largely gas-free star cluster.
Phases (i)-(iii) are encompassed in the birth conditions as deduced by \cite{MK12}. 
All of these processes are contributing to the star cluster's evolutionary path in a complex way as depicted by the bottom solid arrow pointing to the gas-free star cluster.
This point marks the start of our $N$-body simulations, where we assume the embedded pre-gas-expulsion cluster (gray-shaded panel) has expanded and re-virialised after the residual gas has been removed \citep{BK17}. 
Thus, we do not aim to simulate the formation of the cluster but the evolution starts from the expanded gas-free state. 
The long-term evolution of the star cluster and its tidal stream is controlled by the net effect from the stellar evolution, stellar dynamics and the Galactic potential.
In this work, we focus on how the content of massive stars can affect the evolution, which causes the degeneracy of the evolution track and brings the challenge to identify the initial condition from the present-day observational property. 

Observationally, gas-free star clusters are well resembled by Plummer-like profiles \citep{Plummer1911, Roeser+11, Roeser2019a}
and thus our initial set-up is empirically motivated. 
Using the initial coordinate described in Section~\ref{sec:orbit} in the Milky-way potential, we generate a grid of star-cluster models by varying the initial total masses $M_0$ and the initial half-mass radii $R_{\mathrm{h},0}$ (see Table~\ref{tab:mr}). 
We combine each of $M_0$ and $R_{\mathrm{h},0}$, and thus, the total number of model sets is 15.
The reference name of each set is described in Table~\ref{tab:mr}.
To investigate the stochastic effect, it is necessary to carry out a large number of simulations with different random seeds.
Thus, for each set, we generate 300 models by randomly sampling the IMF from \cite{Kroupa2001} with the mass range of $0.08-150~M_\odot$ and the positions and velocities of stars from the Plummer density profile \citep{AHW74}.
The stochastic effect also naturally results in a different initial mass of OB stars ($\mob$), where we assume OB stars have the zero-age mass sequence mass $>8 M_\odot$.


\begin{table}
    \centering
    \caption{The initial parameters of $N$-body models}
    \begin{tabular}{ll}
    \hline
    \noalign{\smallskip}
    $M_0 [M_\odot]$     &  800, 1000, 1200, 1400, 1600\\
    $R_{\mathrm{h},0}[pc]$ &  0.5, 1, 2 \\   
    \noalign{\smallskip}
    \hline
    \end{tabular}
    \tablefoot{The initial total masses $M_0$ and the initial half-mass radii $R_{\mathrm{h},0}$ of the star cluster model sets. The reference name of each set combines the values of $M_0$ and $R_{\mathrm{h},0}$ without the decimal point, such as M800R05, M1000R1, M1600R2.}
    \label{tab:mr}
\end{table}


\section{Results}
\label{sec:result}

\subsection{The variation of $\mob$ due to randomly sampling of IMF}

\begin{figure}[htbp]
    \centering
    \includegraphics[width=\columnwidth]{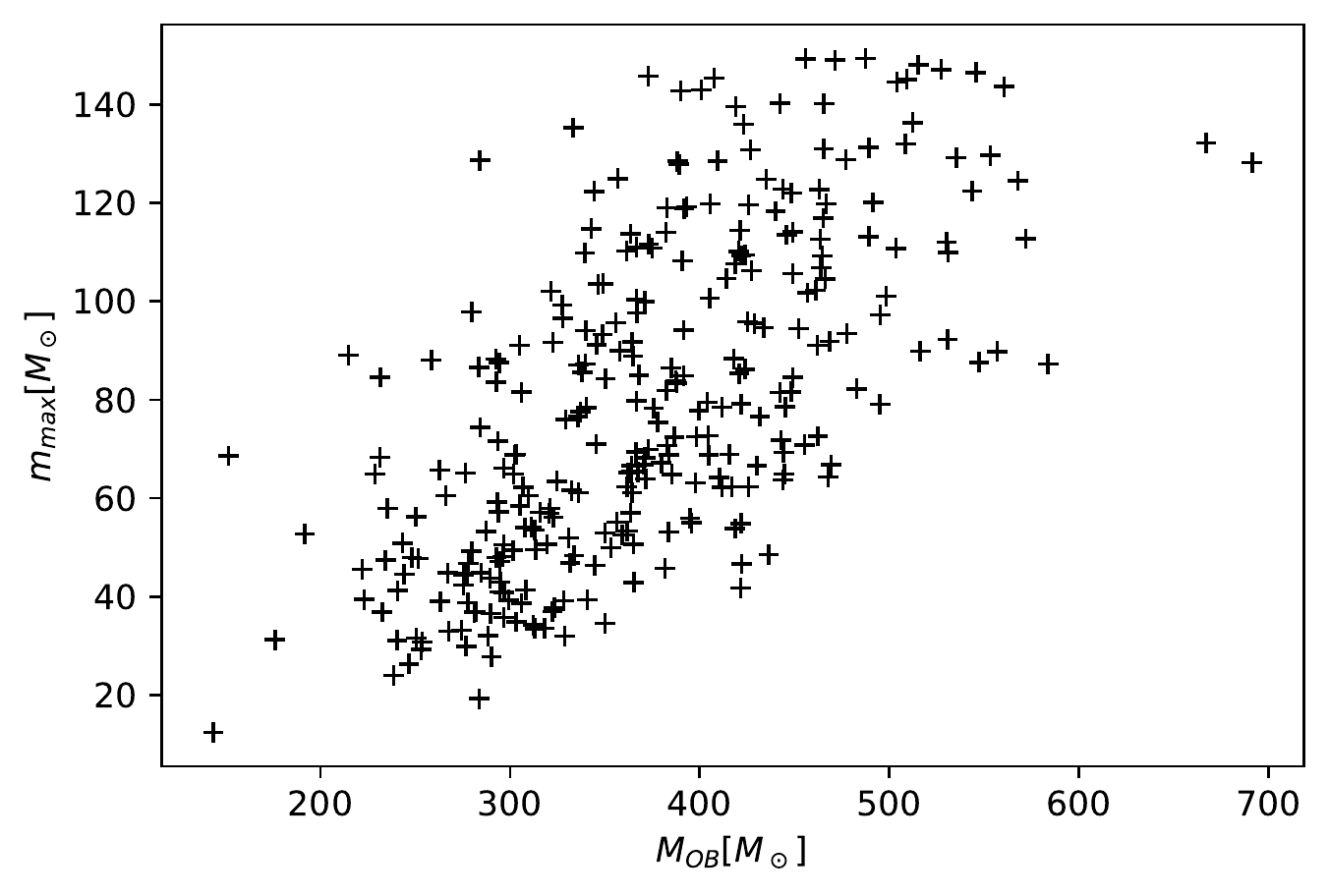}
    \caption{The mass of the heaviest star ($\mmax$) and the total mass of OB stars ($\mob$) by randomly sampling the Kroupa (2001) IMF in each of the 300 Hyades-like models. The initial total mass of each model is $1600~M_\odot$.}
    \label{fig:mmaxmob}%
\end{figure}

The randomly sampling of IMF results in a large dispersion of the mass of the heaviest star ($\mmax$) and $\mob$ model by model.
Fig.~\ref{fig:mmaxmob} shows the $\mmax - \mob$ relation from all 300 models in the M1600R2 set.
$\mob$ varies from  $150~M_\odot$ to $600 M_\odot$ and $\mmax$ varies from 20 $M_\odot$ to  $140~M_\odot$.
The maximum $\mob$ comprises $37.5$ per cent of the initial mass of the cluster. 
Thus, it is expected that the dynamical impact from the OB stars can differ significant model by model.
%


\subsection{The stellar winds from OB stars}

%


\begin{figure*}[htbp]
    \centering
    \includegraphics[width=0.8\textwidth]{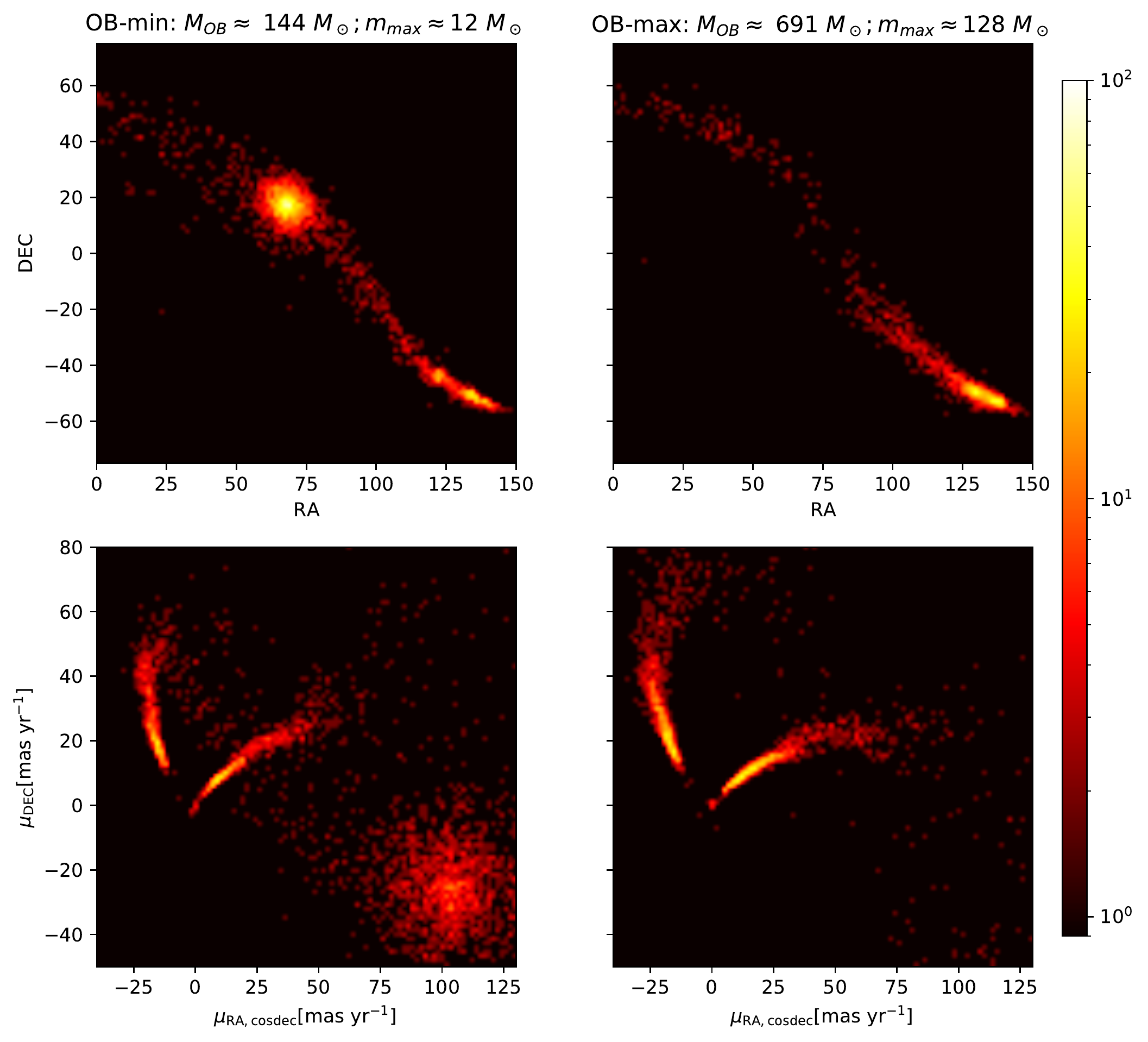}
    \caption{The comparison of the number density map for the positions and proper motions of stars at the age of 648 Myr in the two models with the minimum and the maximum $\mob$ from the M800R2 set (OB-min and OB-max models).}
    \label{fig:comp2}%
\end{figure*}

To investigate how $\mob$ affects the present-day morphology, we pick up the two models with the minimum and the maximum $\mob$ in the M1600R2 set (hereafter named as ``OB-min'' and "OB-max" models, respectively).
Fig.~\ref{fig:comp2} compares the present-day positions and proper motions of stars in the ICRS frame of the two models.
The OB-min model in the left panels has $\mob \approx 154 M_\odot$ and $\mmax \approx 20 M_\odot$, while the OB-max model in the right panel has a much larger $\mob$ and its $\mmax$ is comparable to $\mob$ of the OB-min model.
Their present-day morphology looks completely different: the host star cluster still exists in the OB-min model while it has already been disrupted in the OB-max model.
This suggests the strong impact from $\mob$ on the long-term dynamical evolution. 

The proper motions of the two branches of tidal tails also show different distributions (the upper regions of the lower panels in Fig.~\ref{fig:comp2}). 
This may be caused by the different offsets of the central position and velocity of the two models.
In the OB-min model, we can correct the central coordinate to be the exact observed one.
But it is difficult for the OB-max model where the center cannot be determined.

\begin{figure}[htbp]
    \centering
    \includegraphics[width=0.8\columnwidth]{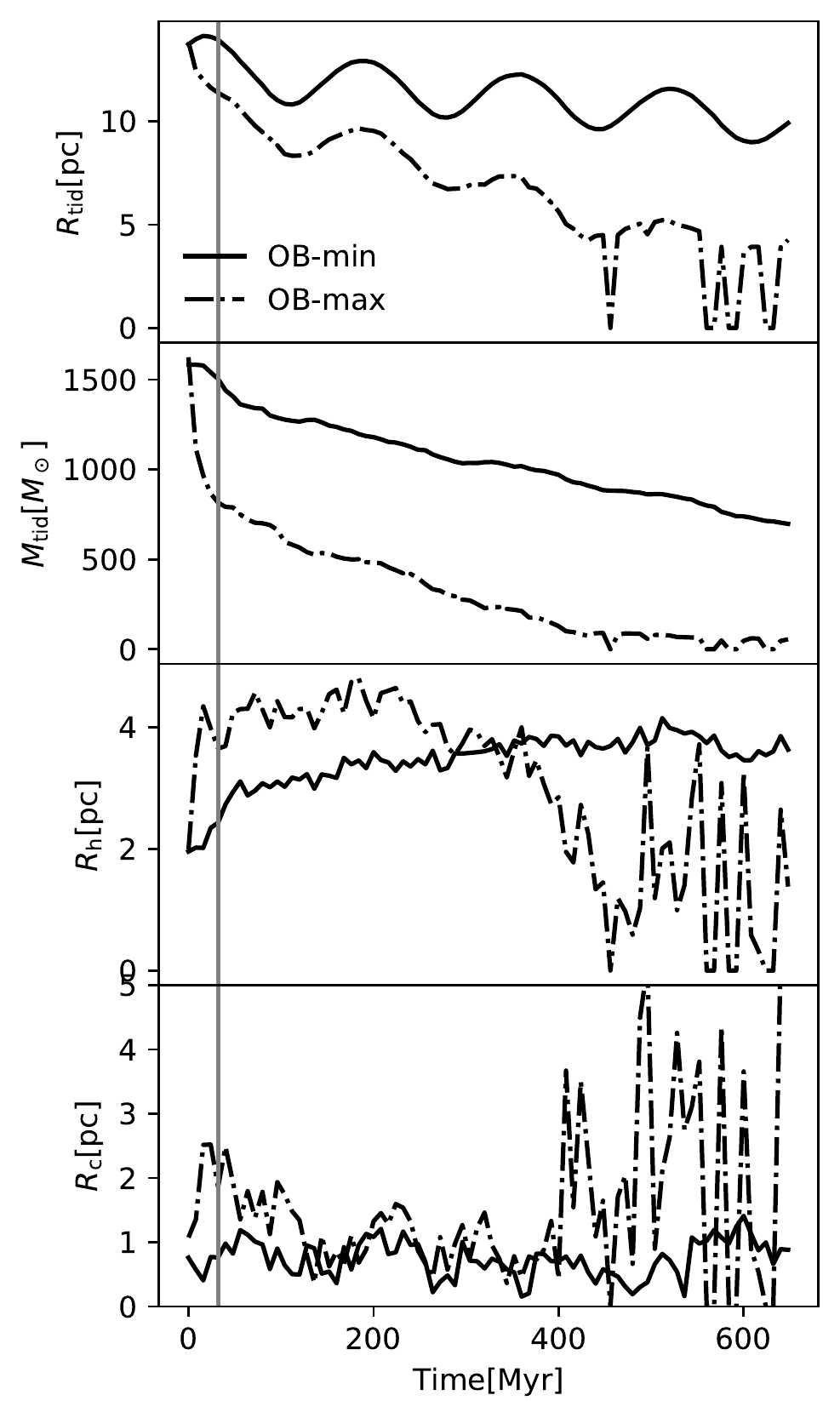}
    \caption{The comparison for the evolution of the global properties for the OB-min and OB-max models shown in Fig.~\ref{fig:comp2}. 
    From the top to the bottom: tidal radii ($\rtid$), the total masses inside $\rtid$ ($\mtid$), the half-mass radii ($\rh$) and the core radii ($\rc$).
    The vertical line refers to 30 Myr.}
    \label{fig:compr}%
\end{figure}

To investigate when the difference started to develop, we compare the evolution of several global parameters for the two models including the tidal radii ($\rtid$), the total masses inside $\rtid$ ($\mtid$), the half-mass radii ($\rh$) and the core radii ($\rc$).
To calculate $\rc$, we apply the method defined by \cite{Casertano1985} as
\begin{equation}
    \rc = \sqrt{\frac{\sum_i \rho_i^2 r_i^{\prime2}}{\sum_i \rho_i^2}},  
    \label{eq:rc}
\end{equation}
where $\rho_i$ is the local density of object $i$ estimated by counting 6 nearest neighbors, and $r_i^{\prime}$ is the distance to the central position of the system.
The center is estimated by the density weighted average:
\begin{equation}
    \mathbf{r}_{\mathrm{cm}}  = \frac{\sum_i \rho_i \mathbf{r}_i}{\sum_I \rho_i}
\end{equation}
where $\mathbf{r}_i$ is the position vector of each star or compact object in the coordinate system used in the simulation.
During the simulation, the origin point of the coordinate system follows the motion of the potential weighted center of the star cluster. 
Notice that the definition of $\rc$ from observation is different. 
Especially, compact objects cannot be detected by observations.
For massive star clusters where many BHs exist, $\rc$ in $N$-body models can be very different from that of observations.
In this study, only a few BHs can form and their impact on the calculation of $\rc$ is neglectable.
We have confirmed this in our models.

In the Galactic potential, there is no precise definition of $\rtid$.
But we can obtain the rough estimation via 
\begin{equation}
    \rtid = R_{\mathrm{gal}} \left ( \frac{\mtid}{3 M_{\mathrm gal}}\right )^{\frac{1}{3}}
\end{equation}
\citep{Binney1987}, where $R_{\mathrm{gal}}$ is the distance to the Galactic center and $M_{\mathrm {gal}}$ is the effective mass of the Galaxy (the mass enclosed by $R_{\mathrm{gal}}$).
In our analysis, we first calculate the (positive) Galactic potential at the cluster center ($P_{\mathrm{gal}}$) and then approximate $M_{\mathrm{gal}}$ via 
\begin{equation}
   M_{\mathrm{gal}} =  \frac{P_{\mathrm{gal}} r_{\mathrm{gal}}}{G}
\end{equation}
where $G$ is gravitational constant.
At the beginning, $\mtid$ is unknown, thus the total mass of all objects is used as the starting $\mtid$.
Then a few iterations are necessary to obtain the consistent $\rtid$ and $\mtid$.
We stop the iteration when the difference between the new and the old $\rtid$ is less then 1 percent.
When the cluster reaches the disruption phase, the iteration may not converge to a positive value of $\rtid$, i.e., $\rtid$ becomes zero.
Once $\rtid$ is determined, the half-mass radii ($\rh$) is calculated by counting all stars inside $\rtid$.

The comparison of these parameters for the two models is shown in Fig.~\ref{fig:compr}. 
The difference already appears during the first 30 Myr. 
$\rtid$ and $\mtid$ of the OB-max model fast decrease due to the strong wind mass loss of OB stars.
More than half of the initial mass has lost during this period.
As the gravitational potential changes, $\rh$ and $\rc$ of the OB-max model also increases significantly at the beginning.
After approximately 400 Myr, $\rh$ decreases while $\rc$ significantly increases.
This indicates that the star cluster loses (virial) equilibrium, and thus, it rapidly expands and approaches to the disruption \citep{Fukushige1995}.
This is also reflected by the oscillation of $\rtid$, where zero appears sometimes. 

In contrast, the OB-min model has a smooth evolution and the star cluster still survive at 648 Myr with more than half of the initial mass being inside $\rtid$.
The evolution of $\rh$ and $\rc$ is flat until $648$ Myr.

\subsection{The general trend depending on $\mob$}

\begin{figure}[htbp]
    \centering
    \includegraphics[width=0.8\columnwidth]{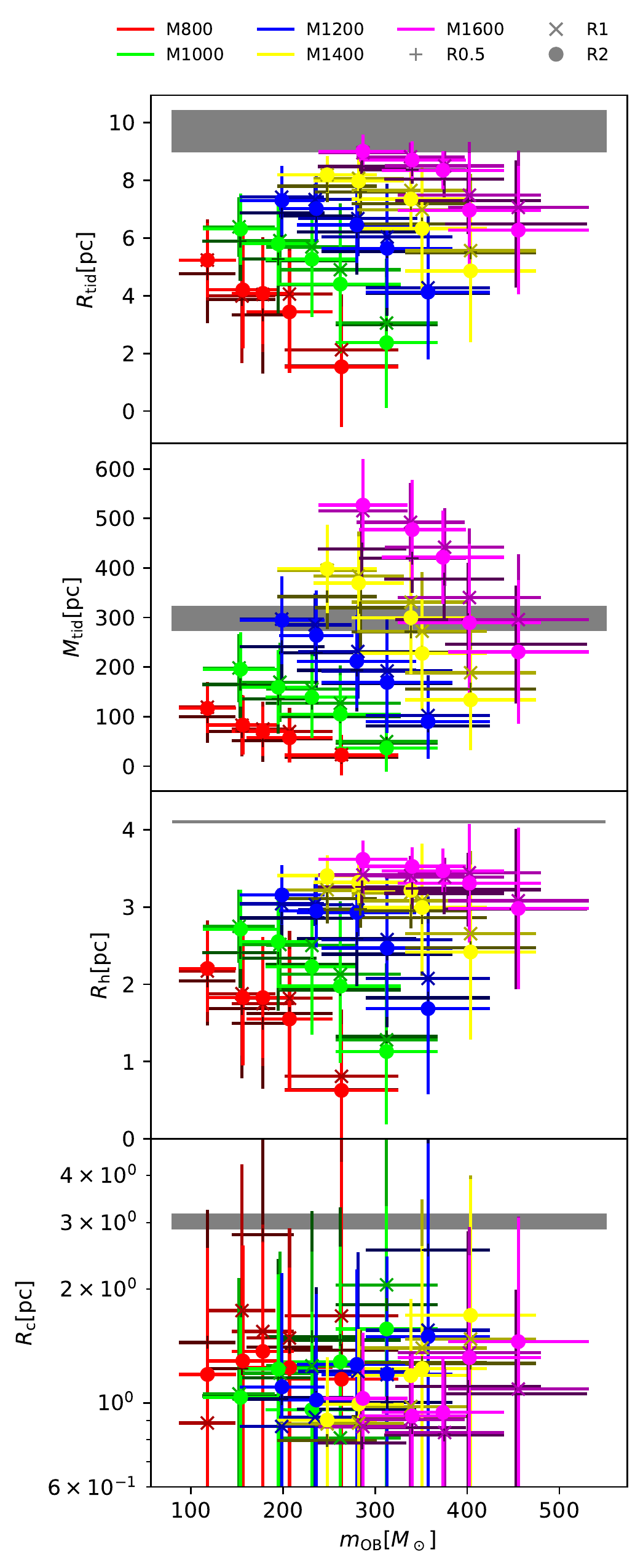}
    \caption{The comparison for the evolution of the global properties for all models with different $M_0$ (represented by colors) and different $R_{\mathrm{h,0}}$ (represented by markers and brightness of colors). From the top to the bottom are: $\rtid$, $\mtid$, $\rh$ and $\rc$. 
    The 300 models in each set are collected into the bins of $\mob$. 
    The error bars represent the standard deviation. 
    The observed values of Hyades are shown as horizontal lines or regions (with uncertainties).}
    \label{fig:mrset}%
\end{figure}

The OB-min and OB-max are two specific cases. 
To investigate the general trend, we separate the 300 models in each model set into five bins of $\mob$.
Each bin contains the same number of models (60). 
Then, for each model in the bin, we obtain $\rtid$ and $\mtid$ at 648 Myr and the average values of $\rh$ and $\rc$ from the snapshots at 632, 640 and 648 Myr. 
The average can reduces the fluctuation that appears in the evolution of $\rh$ and $\rc$ as shown in Fig.~\ref{fig:compr}.
Finally, we calculate the statistic average and the standard deviation of each parameter (including $\mob$) inside each bin.
The result is shown in Fig.~\ref{fig:mrset}.
There is a pronounced trend that a larger $\mob$ results in smaller $\rtid$ and $\mtid$ (faster mass loss).

The reference data from Hyades are shown as grey horizontal lines and regions (with uncertainties). 
Most of models have slightly lower $\rtid$, $\rh$ and $\rc$ compared to those of the observational data.
There is a degeneracy between $\mob$ and $M_0$ that many combinations of the two parameters can result in the similar present-day property of star cluster.
For example, all five $M_0$ set (different colors) can result in the same $\rtid$ ($4-8$~pc), $\mtid$ ($100-300 M_\odot$), $\rh$ ($2-3$~pc) and $\rc$ ($1-2$~pc).
This suggests that it is hard to determine the initial condition by only checking these four parameters.

The impact of $R_{\mathrm{h,0}}$ is not significant. 
Generally, smaller $R_{\mathrm{h,0}}$ result in larger $\rtid$ ($\mtid$). 
But the scatter is large and the effect is less than that from $M_0$ and $\mob$.

\subsection{The long-term dynamical impact from black hole binaries}
\label{sec:bbh}

\begin{figure*}[htbp]
    \centering
    \includegraphics[width=\textwidth]{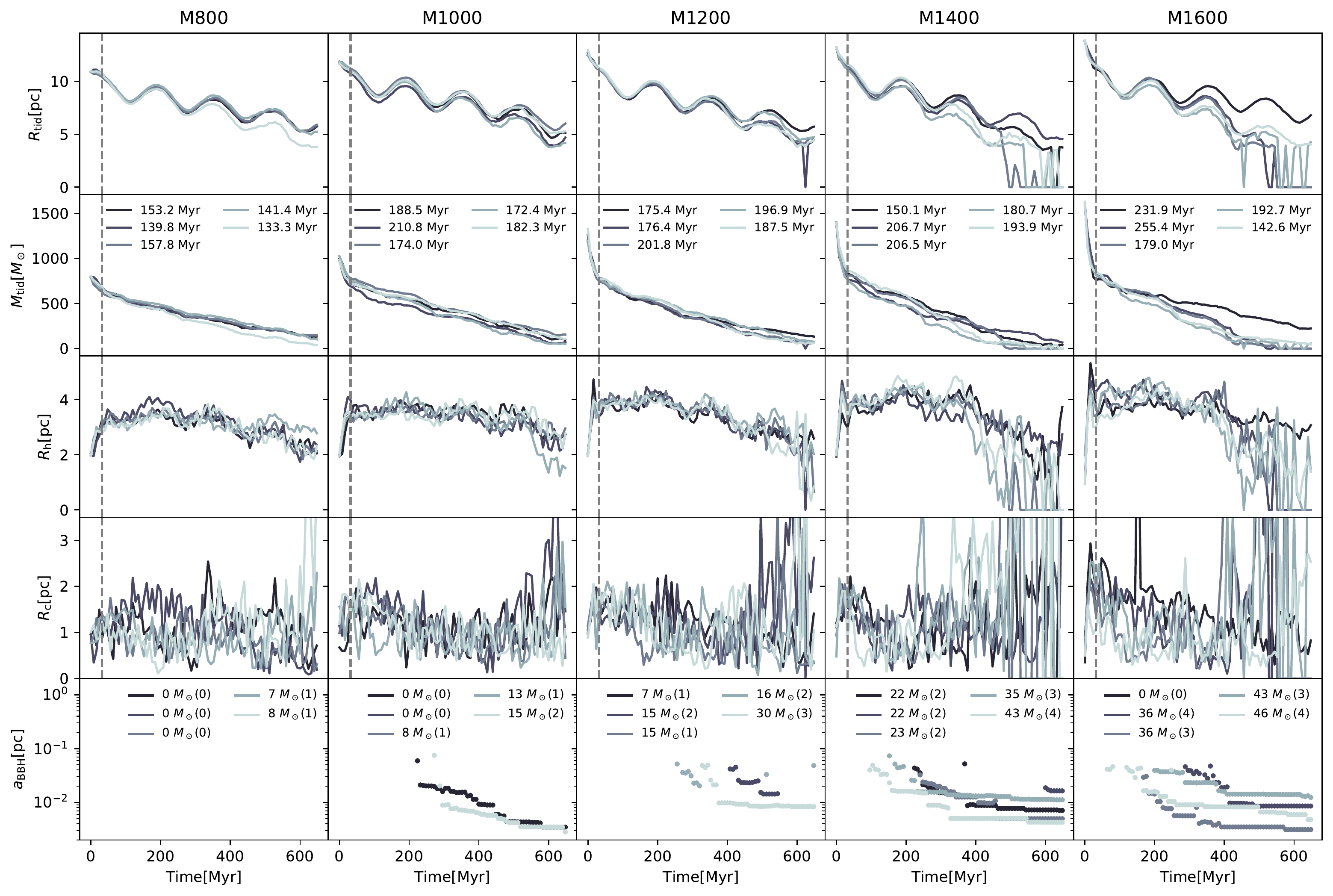}
    \caption{The evolution of the global properties, from the top to the bottom: $\rtid$, $\mtid$, $\rh$, $\rc$ and $\abbh$, for the models with a similar property at 32 Myr to those in the OB-max model. The columns separates the models with different $M_0$. For each $M_0$, a brighter color indicate a higher total mass of BHs ($\mbh$). The half-mass relaxation time ($\trh$) at 32 Myr are shown in the second row of panels. The values of $\mbh$ and the corresponding numbers of BHs (in brackets) are shown in the last row of panels.}
    \label{fig:findsim32}%
\end{figure*}

The major impact from stellar wind mass loss of OB stars appears at the first 32 Myr as shown in Fig.~\ref{fig:compr}.
Thus, if a model has a different initial condition but shows similar properties (e.g. mass, density and $\rh$) at 32 Myr as those in the OB-max model, they may have identical evolution after that.
To investigate this, we measure the difference between models as 
\begin{equation}
    \Delta = \left (\frac{\mtid - M_{\mathrm{tid,ref}}}{M_{\mathrm{tid,ref}}} \right )^2 + \left (\frac{\rh - R_{\mathrm{h,ref}}}{R_{\mathrm{h,ref}}} \right )^2  + \left (\frac{\nrh - N_{\mathrm{rh,ref}}}{N_{\mathrm{rh,ref}}} \right )^2 
    \label{eq:delta}
\end{equation}
where $\nrh$ is the number of stars within $\rh$ and the suffix ``ref'' refers to the parameters of the OB-max model.
The values of $\mtid$ at 32 Myr are used for comparison.
$\rh$ and $\mrh$ have large fluctuations as shown in Fig.~\ref{fig:compr}.
Thus, their averaged values at 24, 32 and 40 Myr are compared.

\cite{Baumgardt2003} showed that the dissolution time of a star cluster depends the strength of tidal field ($\rtid/\rh$) and the relaxation time.
The formula of the half-mass relaxation time for one component system can be described as \citep{Spitzer1987}
\begin{equation}
  \trh \approx 0.138 \frac{N^{1/2} \rh^{3/2}}{m^{1/2} G^{1/2} \ln \Lambda},
  \label{eq:trh}
\end{equation}
where $\Lambda=0.02 N$, $m$ is the mass of star (we use the average mass instead) and $G$ is the gravitational constant.
The factor of $0.02$ follows the measurement from \cite{Giersz1996}.
The three parameters in Equation~\ref{eq:delta} can reflect the tidal effect and $\trh$.

We select models with first five smallest $\Delta$ for each $M_0$ set.
Fig.~\ref{fig:findsim32} shows the comparison for the dynamical evolution of these models.
The models with a larger common $M_0$ tend to have a more divergent long-term evolution.
For the M800 set, all models except one have an identical evolution, while the M1600 models show large differences of $\rtid$ and $\mtid$ after 200 Myr.
A part of M1600 models are still bound at 648 Myr while others have been disrupted.
Such a trend depending on $M_0$ suggests that the stellar wind mass loss at the beginning is not the only reason for the divergent evolution like that of the OB-min and OB-max models.

The values of $\trh$ are shown in the second row of panels in Fig.~\ref{fig:findsim32}.
Generally, the model with a short $\trh$ loses mass faster, but it is not the case for the two models in the M1600 set with $\trh=231.9$~Myr and $255.4$~Myr (hereafter named as R1 and R2 models, respectively).
The R2 model has a large $\trh$ but dissolute faster.
In the meantime, some models in the M1600 set have larger $\trh$ than those of the M800 set, but the former has already been disrupted at the present day while the latter still survives.
Thus, the different evolution cannot be explained by $\trh$.

Instead, we find that the existence of BH binaries (BBHs) play an important role.
The total masses ($\mbh$), the number of BHs at 32 Myr and the evolution of semi-major axies of BBHs are shown in the bottom panels in Fig.~\ref{fig:findsim32}.
The R2 model in the M1600 set has 4 BHs with $\mbh=36 M_\odot$ while the R1 model has no BH.
This can explain why the R2 model loses mass faster.
The decrease of $\abbh$ is found in all models with BBHs in Fig.~\ref{fig:findsim32}, which indicates the dynamical effect of binary heating \citep[e.g.,][]{Binney1987,Spitzer1987,Mackey+08,Breen2013,Wang2020a}.
When light stars have close encounters with the BBH, they gain a large kinetic energy and are easily kicked out from the center of a star cluster while the separation of the BBH shrinks.
Such a process can drive the expansion of the stellar halo and accelerate the escaper generation.
Thus, a star cluster containing more massive BBHs tends to lose mass faster.
In summary, $\mob$ not only affect the strength of stellar wind at the early phase of the star cluster, but also influences the formation of BBHs and the long-term evolution of the system.

\subsection{The structure of tidal streams}

\begin{figure*}[htbp]
    \centering
    \includegraphics[width=\textwidth]{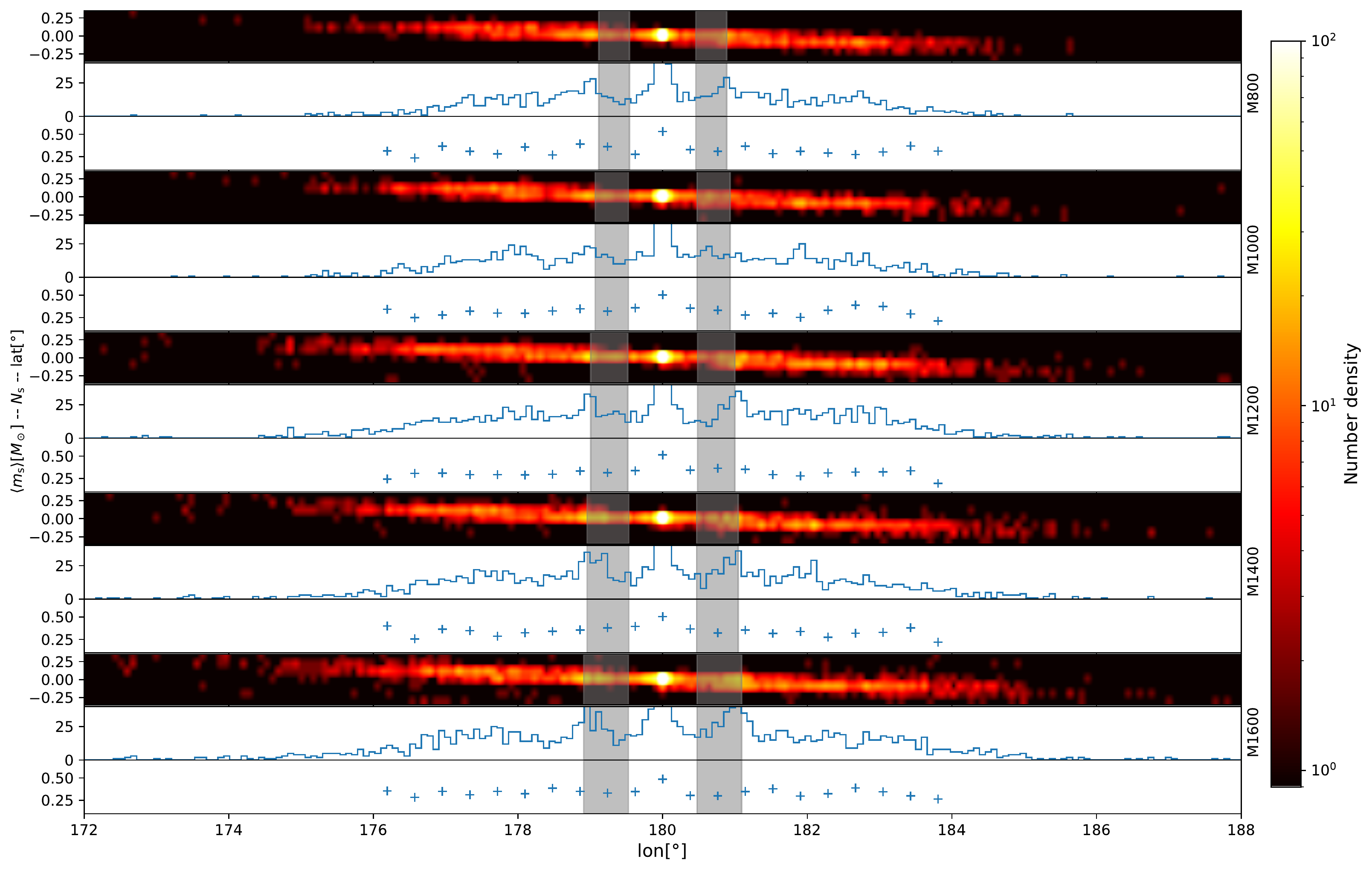}
    \caption{The comparison for the morphology, the density and the average mass of stars for the five models with different $M_0$ and the minimum $\Delta$ referring to the OB-min model in the M800R2 set. For each model, the three sub-panels show the number density map of stars in the spherical coordinate system of the Galactocentric reference frame, the histogram of star counting ($\ns$) and the average mass of stars ($\mave$) along the longitude, respectively. The compacted objects (WDs, NSs and BHs) are excluded.}
   \label{fig:M800R2sim}%
\end{figure*}

It is difficult to reconstruct the initial condition from the present-day property of a star cluster because of the random value of $\mob$.
However, the tidal stream, which records the history of escapers, may provide an additional constraint.
To investigate this, we select the OB-min model in the M800R2 set as the reference, and then, for each $M_0$, we find the most similar model with the minimum $\Delta$ at the present day. 
Totally, five models with different $M_0$ are collected.

The morphology, the number density distribution and the average mass distribution of these (five) models at the present day are shown in Fig.~\ref{fig:M800R2sim}.
For a better comparison, we correct the central position of each model to be located at the observed position of the Hyades.
The morphology (density map) is shown in the Galactocentric reference frame with the spherical coordinate system.
In the longitude--latitude plot, the stream distributes in a regular shape.
Thus, it is easy to count the number of stars along the stream (the longitude) and detect the sub-structures like the overdensity generated by the epicyclic motion \citep{Kuepper2008}.

For all models, the lengths of the streams are similar, and the overdensities appear at the 179~$\deg$ and 181~$\deg$ along the longitude. 
\cite{Kuepper2008} showed that for a point-mass potential with a circular orbit, the distance of the overdensity to the center can be described as
\begin{equation}
    R_{\mathrm{den}} = \rtid \frac{4 \pi \Omega (4 \Omega^2 - \kappa^2)}{\kappa^3}
    \label{eq:overdensity}
\end{equation}
where $\kappa$ and $\Omega$ are epicyclic and circular frequencies, respectively.
Based on the observational data of the solar neighborhood, $\kappa/\Omega \approx 1.35$ \citep{Binney1987}.
Thus, $R_{\mathrm{den}}\approx 3.54 \pi \rtid$.
Since $\rtid$ evolves, we roughly estimate the range of $R_{\mathrm{den}}$ (the grey region in Fig.~\ref{fig:M800R2sim}) by using the minimum and the maximum values of $\rtid$ during the simulation.
To calculate the longitude of $R_{\mathrm{den}}$, $R_{\mathrm{gal}}$ is needed.
From Figure~\ref{fig:orbit}, $R_{\mathrm{gal}}$ varies between $7$ to $9$ kpc. 
We simply use the present-day value to obtain a rough estimation.
The predicted positions of the overdensities are consistent with those of the models. 
Since $R_{\mathrm{den}}$ is independent of $M_0$, the morphology of the tidal streams does not show a pronounced difference.

We also analyze the average mass ($\mave$) along the longitude. 
There is a clear sign of mass segregation that $\mave$ is much larger inside the star cluster compared to that along the stream.
But there is also no pronounced difference depending on $M_0$.

Although the structures and the mass functions of the five tidal streams are similar, we find that the star counts inside and outside $\rtid$ strongly depend on $M_0$. 
Table~\ref{tab:M800R2sim} lists the total masses and the numbers of stars with apparent G-band magnitudes $<20$ inside and outside $\rtid$ for these 5 models, respectively.
All models have a similar $\mtid$ and $\ntid$ while the low-$M_0$ model has a small $\mout$ and $\nout$.
As the high-$M_0$ model has a large $\mob$, the dissolution of the star cluster is faster because of stronger stellar wind mass loss and dynamical heating from BBHs. 
Thus, the fraction of the number of stars inside $\rtid$ ($\ftid = \ntid/N$ where $N$ is the total number of stars) can help to estimate $M_0$ and $\mob$ for an observed star cluster.

From the Gaia EDR3 catalogue, 862 objects with a similar magnitude cutoff were detected, 293 of which belong to the leading tail and 166 to the trailing tail. 
All $\nout$ in Table~\ref{tab:M800R2sim} are significantly higher than the observed counts.
But it is difficult to make a solid conclusion based on the direct comparison of the $N$-body model to the observation.
For Gaia data, a strict selection is needed to remove contamination and obtain signal in the tails, which affects completeness. 
With a looser constraint, the contamination is higher.
In addition, an unresolved binary is counted as one object and large fraction of these are filtered out by quality cuts applied to the Gaia data \citep{Jerabkova2021}.
Due to the overlap of objects along the line-of-sight and a higher fraction of binaries inside the star cluster, the numbers of stars might be underestimated.
Meanwhile, the background contamination is expected to be higher along the tails.
A better solution is to use the $N$-body model as the reference to create mock observations considering the Gaia uncertainty and background contamination.
Then compare the model with the observation to obtain a better constraint. 
For such a purpose, the assumption of primordial binaries is also necessary.
This will be our future effort.

This result suggests one strategy to constrain the initial condition of a star cluster. 
First, we find a group of models with different $M_0$ and $\mob$ that can result in the same $\rtid$, $\mtid$ and $\rh$. Then, we use $\ftid$ to determine which combination of $M_0$ and $\mob$ best fits the observational data.
Because of the high performance of \textsc{petar}, it is practically possible to generate a grid of star clusters models to achieve this goal.

\begin{table}
    \centering
    \caption{The parameters of the model set shown in Fig.~\ref{fig:M800R2sim}}
    \begin{tabular}{cccccc}
    \hline
    \noalign{\smallskip}
    Mode set & $\mob$ &$\mtid$ & $\mout$ & $\ntid$ & $\nout$  \\
    \hline
    \noalign{\smallskip}
    M800R2   & 27.0  & 196.3 & 471.9 &  186 & 735 \\
    M1000R1  & 207.2 & 196.5 & 518.2 &  184 & 761 \\
    M1200R05 & 165.5 & 196.1 & 658.6 &  173 & 958 \\
    M1400R05 & 348.8 & 198.5 & 727.1 &  202 & 973 \\
    M1600R05 & 456.9 & 196.2 & 857.0 &  180 & 1208 \\
    \hline
    \noalign{\smallskip}
    \end{tabular}
    \tablefoot{The corresponding model set, the initial OB star mass ($\mob$), the total mass inside $\rtid$ ($\mtid$) and outside $\rtid$ ($\mout$), the number counts for stars with apparent G-band magnitude $<20$ inside $\rtid$ ($\ntid$) and outside $\rtid$ ($\nout$) for the five models shown in Fig.~\ref{fig:M800R2sim}. Their $\rtid\approx 6.5$~pc and $\rh \approx 3.0$~pc at 648 Myr. The mass unit is $M_\odot$.}
    \label{tab:M800R2sim}
\end{table}



\section{Discussion and conclusion}
\label{sec:conclusion}

In this work, we investigate how the content of OB stars affects the long-term evolution of Hyades-like open star clusters and their tidal streams. 
The models are designed to have the initial condition, including the shape of the IMF and the density profile, to be identical. The random number of OB stars can result in  a significantly different dynamical evolution of the system.
As the example in Fig.~\ref{fig:comp2} shows, the OB-min model with a small $\mob$ can survive as a star cluster until the present day, while the OB-max model with a large $\mob$ has already dissolved 200 Myr ago. 

The stellar wind mass loss of OB stars during the first 32 Myr and the BBH dynamical heating both accelerate the mass loss of star clusters (Fig.~\ref{fig:findsim32}) and cause the difference.
Thus, assuming the stochastic variation of $\mob$ being the correct physical model, it is not possible to constrain the initial condition (e.g. $M_0$ and $\mob$) of an observed star cluster confidently by only checking the present-day properties inside $\rtid$ (e.g., $\mtid$, $\rh$ and $\rc$).
Meanwhile, we find that the morphology (overdensity) and average mass distribution along the tidal streams are independent on $M_0$ (see Fig.~\ref{fig:M800R2sim}).
However, the star counts inside and outside $\rtid$ ($\ftid$; Table~\ref{tab:M800R2sim}) can help to remove the degeneracy and help to constrain the initial condition. 

There are a few aspects that are not included in our analysis.
We did not consider the influence from the uncertainty of the ages in the observation. 
We expect that the age may affect the determination of $M_0$ by using $\ftid$, but it may also influence the length of the tidal stream and the position of the overdensity.
Thus, we may disentangle the impacts from the age and the $\mob$.

There are no primordial binaries in our models. 
Observations have shown that massive stars in open clusters are most likely all in multiple systems \citep{Sana2012,Duchene2013,Moe2017}.
The dynamical interactions of these multiple systems can result in ejections and mergers \citep[e.g.][]{Wang2019}.
As a result, the numbers of retained OB stars and formed BHs decrease.
In addition, the primordial binaries can also attend the binary heating at the early phase that affects the timescale of core collapse \citep[e.g.][]{Heggie2006}.
All of these subsequently affect the long-term dynamical evolution of clusters.
In the future work, we will also consider the impact from primordial binaries.

The gas expulsion  process can also significantly affect the dynamics of the star cluster during the gas embedded phase \citep[e.g.,][]{Wang2019,Fujii2021a,Fujii2021b} and the formation of tidal streams \citep{Dinnbier2020a, Dinnbier2020b}.
The UV radiation, stellar winds and supernovae from OB stars all can drive the gas expulsion. 
Thus, the gas expulsion is also sensitive to the stochastic nature  of $\mob$.
A stronger feedback because of more OB stars may quench the star formation earlier, and thus, the star formation efficiency becomes lower.
As a result, the damage to the stellar system is also stronger and a larger number of escapers is expected to appear during the first few Myr. 
By considering gas expulsion, we expect an even larger scatter of the evolution tracks due to the variation of $\mob$. 

As a caveat it is noted that
we applied a simple Milky Way potential (MWPotential2014). There is no time-dependent evolution of the potential, no bar and no spiral arm. 
Since we are not aiming at obtaining the precise morphology of the tidal stream for the Hyades,  missing these components does not significantly influence the major conclusion.
Since \textsc{galpy} supports to build up a more realistic Galactic potential, we will investigate such an effect in the future work.

We assumed in this work that the formation of OB stars is purely statistical implying a large variation of $\mob$.
But if the star formation has some degree of self-regulation such that the total number and mass of OB stars depends on the initial mass of the cluster-forming gas cloud, the scatter of $\mob$ would be much smaller than the case studied here.
Then, constraining the initial conditions of an observed open cluster would be much easier.
For example, if the star formation is highly self-regulated such that the stars formed can be described by optimally sampling the IMF \citep{Kroupa+13}, $\mob \approx 289.81 M_\odot$ and $\mmax \approx 53 M_\odot$ for $M_0=1600~M_\odot$ and only a tiny scatter of $\mob$ is allowed.
Therefore, by invoking stochastic sampling from the IMF, our work provides the upper limit on the uncertainty in constraining the initial conditions of a given open star cluster. 

\begin{acknowledgements}
L.W. thanks the financial support from JSPS International Research Fellow (School of Science, The university of Tokyo). We thank Jo Bovy for the help on the implementation of \textsc{galpy} interface in \textsc{petar} and Qi Shu for the help on the debugging of the code. TJ  acknowledges  support  through  the  European  Space  Agency  fellowship  programme.
\end{acknowledgements}

%
%

\begin{appendix} 
\section{\textsc{petar} - \textsc{galpy} }
\label{sec:galpy}

\textsc{galpy} \citep{Bovy2015} is the state-of-the-art code to modelling the orbits of stars in different kinds of galactic potentials. 
The online document provides rich information about the usage of the code\footnote{https://docs.galpy.org/en/v1.7.0/}.
The code is written mainly in the Python programme language while it also provides the c programme language for coupling to the $N$-body codes with a high computing performance.
A interface has been implemented in \textsc{petar} so that it is possible to trace the formation and evolution of tidal streams.
Table~\ref{tab:pot} lists all names of \textsc{galpy} potentials (version 1.7.0) that can be accessed by \textsc{petar}.
The MWPotential and MWPotential2014 are wrapped potentials \citep[see ][]{Bovy2015}.
The \textsc{petar.galpy.help} tool provides the basic description of each potential type.
It is recommended to use the Python interface of \textsc{galpy} to configure the potential, and then, save the parameters and use petar to read them to initialize the potential in $N$-body simulations.

\begin{table}
    \centering
    \caption{The supported potentials in the \textsc{petar-galpy} interface}
    \begin{tabular}{ll}
Burkert & DehnenBar \\
DehnenCoreSpherical & DehnenSpherical \\
DiskSCF & DoubleExponentialDisk \\
FlattenedPower & Hernquist \\
HomogeneousSphere & Isochrone \\
Jaffe & Kepler \\
King & KuzminDisk \\
KuzminKutuzovStaeckel & LogarithmicHalo \\
MN3ExponentialDisk & MWPotential \\
MWPotential2014 & MiyamotoNagai \\
NFW & PerfectEllipsoid \\
Plummer & PowerSpherical \\
PowerSpherical & PowerTriaxial\\
PseudoIsothermal & SCF \\
SoftenedNeedleBar & SpiralArms \\
TriaxialGaussian & TriaxialHernquist \\
TriaxialJaffe & TriaxialNFW   \\
    \end{tabular}
    \label{tab:pot}
\end{table}

When \textsc{galpy} is switched on, \textsc{petar} uses the Galactic center as the reference frame  without rotation.
To keep the high digital precision for positions and velocities of stars, the  coordinate origin of the stellar system is not the Galactic center but follows the motion of the center of the cluster.
The position and velocity of center referring to the Galactic center are saved as the offset values in the header of the snapshots. 
Using these offset values, users can easily obtain the coordinate frame referring to the Galactic center.
The default way to determine the center is to use the long-range potential weighted average of positions and velocities of all objects.
This may not be the best way.
It is challenging to determine the center when the star cluster is close to the complete disruption. 
Thus, users may need to redetermine the center in the late phase of the evolution by using the snapshots.

The \textsc{petar-galpy} interface supports to add two types of potentials: the co-moving potentials and the fixed central potentials.
The former co-moves with the center of the stellar system.
The latter has a fixed position referring to the Galactic center.
Either type can combine arbitrary time-dependent potentials listed in Table~\ref{tab:pot}.
Thus, it is flexible to study a variety topics.
For example, users can setup a star cluster move in the Milky-way potential, with a gas halo or a dark matter halo surrounding the cluster. 
The gas potential can be reduced by time to represent the gas expulsion.
The details to setup the potentials can be found in the online README of \textsc{petar}\footnote{GitHub page: https://github.com/lwang-astro/PeTar} and from the commander 'petar -h'.

\section{Comparison with observation}
\label{sec:obs}
To construct a bridge to connect the $N$-body simulations and observations, we have implemented a group of data analysis tool written in Python3 (see online manual for details),
Here we briefly describe the functions of this tool.

\subsection{The transformation of reference frames and coordinate systems}

\textsc{petar} equipped with \textsc{galpy} uses the Galactocentric reference frame and the Cartesian coordinate system in the simulation. 
The result cannot be directly compared with the observational data.
The \textsc{petar} data analysis tool includes an interface to transform the snapshot from a $N$-body simulation to the data type of \textsc{astropy.coordinates.skycoord} \citep{astropy2013,astropy2018}.
\textsc{skycoord} is a flexible data type that supports an easy way to transform reference freams and coordinates systems for a group of data with 3D positions and velocities.
Fig.~\ref{fig:comp2} and \ref{fig:M800R2sim} show the snapshot of $N$-body simulations in the ICRS frame (RA, Dec, proper motions) and in the Galactocentric frame (longitude and lattitude), respectively.
These can be easily generated by using this interface.

\subsection{Convergent point check for proper motions}
One of the very useful way of identification of (nearly) co-moving stars in the Gaia catalog, that is using only proper motions and not the mostly absent radial velocity values, is so-called convergent point method. Originally, the convergent point (CP) method has been used to constrain distance to nearby star-clusters 
 \citep[][and references therein]{Smart1939,Jos1999,vanLeeuwen09}.
 The innovative usage of this method to identify co-moving stars instead has been first applied by \citep{Roeser2019b} and later discussed in detail by  \citep{Jerabkova2021}.
The CP method is correcting measured proper motions for on-the-sky projection effects that can be substantial for close-by extended objects. For more details we refer reader to \cite{Jerabkova2021, vanLeeuwen09, Roeser2019b}. 

The mentioned PeTar script computed the CP diagram and allows thus compare simulations with the Gaia and offers thus more possiblities how to search for co-moving stars in the Gaia catalog.

\subsection{Mock observational errors}
In the current version of the tool it is possible to generate expected Gaia EDR3 uncertainties based on stellar magnitudes computed based on respective stellar mass. The detail description will be provided in the upcoming work with demonstration cases. 

%

\end{appendix}

\end{document}